\documentclass[twocolumn,nolinenumbers]{aastex631}
\usepackage{graphicx}
\usepackage{dcolumn}
\usepackage{bm}
\usepackage{amssymb}
\usepackage{amsmath}
\usepackage{soul, color, xcolor}

\usepackage{epstopdf}
\begin{document}

\title{Electromagnetic Flares Associated with Gravitational Waves from Binary Black Hole Mergers in AGN Accretion Disks}

\author[0009-0007-0717-3667]{Zhi-Peng Ma}
\affiliation{Department of Astronomy, School of Physics, Huazhong University of Science and Technology, Wuhan, Hubei 430074, China}\email{kaiwang@hust.edu.cn}

\author[0000-0003-4976-4098]{Kai Wang}
\affiliation{Department of Astronomy, School of Physics, Huazhong University of Science and Technology, Wuhan, Hubei 430074, China}

\author[0000-0003-4773-4987]{Qingwen Wu}
\affiliation{Department of Astronomy, School of Physics, Huazhong University of Science and Technology, Wuhan, Hubei 430074, China}

\author[0000-0001-9449-9268]{Jian-Min Wang}
\affiliation{Key Laboratory for Particle Astrophysics, Institute of High Energy Physics, Chinese Academy of Sciences, 19B Yuquan Road, Beijing 100049, China}
\affiliation{School of Astronomy and Space Sciences, University of Chinese Academy of Sciences, 19A Yuquan Road, Beijing 100049, China}
\affiliation{School of Physics, University of Chinese Academy of Sciences, 19A Yuquan Road, Beijing 100049, China}

\begin{abstract}
The gravitational wave (GW) event GW190521, likely originating from a binary black hole (BBH) merger within an active galactic nucleus (AGN) disk, is associated with the optical flare ZTF19abanrhr. The remnant BHs from BBH mergers can launch the jet and outflow and then interact with the disk medium, which can be responsible for the associated electromagnetic radiations.
In this \textit{letter}, we examine the shock breakout and subsequent cooling emissions from four potential components: the outflow, jet head, jet cocoon, and disk cocoon, all driven by the remnant BH within the AGN disk. Using dynamic models and observational constraints, for GW190521, we identify the parameter space for each component and conclude that either the outflow or the disk cocoon could produce the observed electromagnetic signal, with the disk cocoon requiring more extreme parameters. We present best-fit light curves and spectral energy distributions (SEDs) for both components, showing peak emissions in the UV band for the outflow and spanning optical to UV for the disk cocoon.

\end{abstract}
\keywords{ High energy astrophysics --- Active galactic nuclei --- Gravitational waves --- Black holes}

\section{Introduction} 
\label{sec_intro}
GW190521, a gravitational wave (GW) burst with a luminosity distance of approximately 3931 Mpc, is classified as a binary black hole (BBH) merger with a total mass of around 150 \(M_{\odot}\)~\citep{graham2020candidate}. This very high-mass stellar BBH merger suggests that the merger is unlikely to be an isolated binary and is instead embedded in an AGN disk \citep{cheng1999formation,artymowicz1993star,collin1999star,bartos2017rapid,stone2017assisted,tagawa2020formation,cantiello2021stellar,li2022long,delaurentiis2023gas}. The GW event is associated with the optical flare ZTF19abanrhr \citep{graham2020candidate,ashton2021current,morton2023gw190521}, which begins $\sim 30$ days after the GW detection, rises to its peak over approximately 10 days and then declines over about 60 days, with peak magnitudes around 18.7 in the g band and 18.5 in the r band. Additionally, seven AGN flares have been statistically linked to LIGO/Virgo BBH merger events \citep{graham2023light}.

Isolated BBH mergers in the common environment are believed to not produce accompanying electromagnetic (EM) radiations due to lack of accretion materials, while for BBH mergers embedded in AGN disks, high-energy neutrinos~\citep{zhou2023high,ma2024high} and EM counterparts are expected by accreting disk materials and then interacting with the disk medium. In terms of EM counterparts, some studies have investigated EM emissions focusing on interactions between the disk gas and the jet launched by the remnant BH~\citep{2023ApJ...958L..40W,rodriguez2024optical,tagawa2024shock,chen2024electromagnetic}. In addition to jet-disk interactions, the BBH mergers and remnant BH may power a fast outflow. The shocks formed by the interactions between the outflows and the disk materials are also potential sources of EM emissions. Compared to the jets of merged BHs, the sub-relativistic outflow driven by the merged BH is quasiisotropic, without beaming effects, leading to a more promising detection of GW-EM associations for the currently small number of detected BBH mergers. Several studies have discussed the role of outflows~\citep{wang2021accretiona,wang2021accretionb,kimura2021outflow,chen2023role}.

Currently, studies on EM counterparts associated with BBH mergers are still inadequate, especially for their light curves. To determine the mechanism responsible for the EM signals associated with BBH mergers, in this \textit{letter}, we investigate the light curves of four plausible EM components during interactions of disk medium with the jets and outflow powered by the remnant BHs from BBH mergers, including the jet head, the disk cocoon and the jet cocoon from the jet-disk interaction, and the outflow-disk interaction. Then we apply in the specific case of GW190521 and attempt to explain the observed light curve. Based on observations, we explore the allowed parameter space for these components, the resulting light curves, and SEDs for GW190521.

\section{Outflow and Jet Components Powered by Remnant BH}
\label{sec_out}

We describe the disk midplane through the method proposed by \cite{cantiello2021stellar}, 
which modified the conventional $\alpha$-disk model by considering the gravity normal to the disk plane, momentum transfer, and possible thermal and momentum feedback from newly born/captured stars within the disk. The density profile in the vertical direction has the Gaussian form~\citep{kato1998black,zhu2021high}
\begin{equation}
    \rho_{\rm disk}(r,h)=\rho_0(r){\rm exp}\left(-\frac{h^2}{2H^2}\right),
\end{equation}
where $\rho_0(r)$ is the gas density in the midplane of the disk, $ r$ is the distance to the central SMBH, which can be expressed as a multiple of the SMBH Schwarzschild radius $r=f_{\rm r}R_{\rm Sch}$, $h$ is the vertical distance to the midplane, where $H$ is the disk scale height. The pressure and temperature are $P_{\rm disk}(r,h)=\rho_{\rm disk}/m_pk_bT_{\rm disk}$ and $T_{\rm disk}(r,h)=T_0(r)$, where $m_{\rm p}$ is the proton mass, $k_b$ is the Boltzmann constant and $T_0$ is the temperature in the midplane.

After being kicked into a gas-abundant environment in the disk, the remnant black hole would accrete the disk gas in the way of Bondi-Hoyle-Littleton (BHL) accretion and thus power a fast outflow. The accretion rate is~\citep{bondi1952spherically,hoyle1939effect,edgar2004review}
\begin{align}
\dot{M}_{\rm BHL} &\simeq\frac{4\pi G^2M^2_{\rm bh}\rho_0}{v^3_{\rm k}}\nonumber\\
&\simeq 2.2\times10^{26}{\rm g \ s^{-1}}\left(\frac{M_{\rm bh}}{10^2M_{\odot}}\right)^2\nonumber\\
&\left(\frac{\rho_0}{10^{-10}{\rm g \ cm^3}}\right)
\left(\frac{v_{\rm k}}{10^{7}{\rm cm \ s^{-1}} }\right)^{-3},
\end{align}
where $M_{\rm bh}$ is the mass of remnant black hole, $v_{\rm k}\sim 10^2~\rm km/s$ is the kick velocity~\citep{campanelli2007maximum,gonzalez2007maximum,herrmann2007unequal}. The outflow kinetic luminosity comes from the accretion power
\begin{align}
    &L_{\rm w}=\eta_{\rm w}\dot{M}_{\rm BHL}v^2_{\rm w}\nonumber\\
    =10^{45}{\rm erg/s} &\left(\frac{\eta_{\rm w}}{10^{-1}}\right)\left(\frac{\dot{M}_{\rm BHL}}{10^{26}{\rm g \ s^{-1}}}\right)\left(\frac{v_{\rm w}}{10^{10}{\rm cm \ s^{-1}}}\right)^2,
\label{eq_lum}
\end{align}
where $\eta_{\rm w}$ is the ratio of outflow kinetic luminosity to the BHL accretion power and $v_{\rm w}$ is the velocity of the outflow. The outflow velocity in the BH-disk system can range from several hundred kilometers per second to ultrafast speeds $\sim 0.3 \rm c$ (c is the light speed) driven by either SMBHs~\citep{2003MNRAS.345..705P,2005A&A...431..111B} or microquasars~\citep{2002LNP...589..101F}, even up to relativistic speed~\citep{10.1093/mnrasl/slad182}. Here, we adopt $v_{\rm w}=10^{10}~\rm cm/s$.

When the fast, isotropic outflow encounters the disk gas, a two-shock system is formed. 
In this work, we focus on the evolution of the forward shock, whose radius $R_{\rm s}$ and velocity $v_{\rm s}$ are determined by a set of equations in the thin-shell approximation~\citep{weaver1977interstellar,faucher2012physics,wang2019transient,murase2024interacting}. The related equations can be written as (also see~\cite{ma2024high})

\begin{equation}
\label{eq_mov}
\frac{d(M_{\rm s}v_{\rm s})}{dt}=4\pi R^2_{\rm s}(P_{\rm in}-P_{\rm disk})-\frac{GM_{\rm tot}M_{\rm s}}{R^2_{\rm s}},
\end{equation}

\begin{equation}
\label{eq_eng}
     \frac{dE_{\rm in}}{dt}=\frac{1}{2}\dot{M}_{\rm w}(1-\frac{v_{\rm rs}}{v_{\rm w}})\left[v^2_{\rm w}-\left(\frac{R_{\rm s}}{R_{\rm rs}}\right)^2v_s\right]-4\pi R^2_{\rm s}P_{\rm in}v_{\rm s}.
\end{equation}

\begin{equation}
\label{eq_5}
    v_{\rm rs}=\frac{4}{3}\left(\frac{R_{\rm s}}{R_{\rm rs}}\right)^2v_{\rm s}-\frac{1}{3}v_{\rm w},
\end{equation}

\begin{equation}
\label{eq_pe}
    E_{\rm in}=2\pi P_{\rm in}(R^3_{\rm s}-R^3_{\rm rs}),
\end{equation}

\begin{equation}
    M_{\rm s}=\int_{0}^{R_{\rm s}}{4\pi R^2\rho_{\rm disk}(r, R)dR},
\end{equation}

\begin{equation}
\label{eq_mt}
    M_{\rm tot}=M_{\rm bh}+M_{\rm s}/2.
\end{equation}
Eq.~(\ref{eq_mov}) is the motion equation of the forward shock, where the first and second terms on the right hand represent the pressure and gravity exerted by the total gravitational mass $M_{\rm t}$ respectively, $M_{\rm s}$ is the mass of the shocked gas, $P_{\rm in}$ is the pressure of the shocked outflow, whose value is determined by Eq.~(\ref{eq_eng}) and energy-pressure relation Eq.~(\ref{eq_pe}). Eq.~(\ref{eq_eng}) describes the internal energy evolution of the shocked outflow, where the first and second terms on the right hand represent the internal energy injection rate and the energy loss due to expansion respectively, where the outflow injection rate is $\dot{M}_{\rm w}=2L_{\rm w}/v^2_{\rm w}$. The motion for the reverse shock is given by Eq.~(\ref{eq_5}), which can be derived from mass conservation and Rankine–Hugoniot jump relation~\citep{liu2018can,dermer2009high}.

By solving Eq.~(\ref{eq_mov}) to Eq.~(\ref{eq_mt}), one can get the evolution of $v_{\rm s}$ and $R_{\rm s}$, we can also estimate the thermal luminosity $L_{\rm th}$ of the forward shock, which can be written as~\citep{liu2018can} 
\begin{align}
   L_{\rm th,w}&\simeq\frac{9}{8}\pi R^2_{\rm s}\rho_{\rm s}v^3_{\rm s}\nonumber\\
   &\simeq1.35\times10^{44}{\rm erg \ s^{-1}}\left(\frac{R_{\rm s}}{10^{13} {\rm cm}}\right)^2\nonumber\\
   &\left(\frac{\rho_{\rm disk}}{10^{-10}{\rm g \ cm^{-3}}}\right)^2\left(\frac{v_{\rm s}}{10^{9} {\rm cm \ s^{-1}}}\right)^3,
\end{align}
where $\rho_{\rm s}=4\rho_{\rm disk}$ is the density of shocked gas.
The internal energy carried by the shocked gas until time t is
\begin{equation}
    E_{\rm th,w}(t)=\int_{0}^{t}{L_{\rm th,w}dt}
\end{equation}

In addition to the outflow, the remnant BH may launch a jet through BHL accretion. The jet power can be described similarly to the luminosity in Eq.~(\ref{eq_lum}):
\begin{align}
    &L_{\rm j}=\eta_{\rm j}\dot{M}_{\rm BHL}c^2\nonumber\\
    \simeq10^{46}{\rm erg/s} &\left(\frac{\eta_{\rm j}}{10^{-1}}\right)\left(\frac{\dot{M}_{\rm BHL}}{10^{26}{\rm g \ s^{-1}}}\right),
\end{align}
As the jet propagates, it decelerates and dissipates energy into a lateral cocoon. The jet head, disk cocoon, and jet cocoon may contribute to EM emissions. The velocity of the jet head can be estimated as
\begin{equation}
\label{eq_vh}
    \beta_{\rm h}=\frac{\beta_{\rm j}}{1+\Tilde{L}},
\end{equation}
where $\beta_{\rm j}=v_{\rm j}/c \simeq 1$ with the jet velocity $v_{\rm j}$, and $\Tilde{L}$ is the collimation parameter. The evolution of the location and velocity of the jet head can be obtained using Eq.~(\ref{eq_vh}). The kinetic luminosity of the jet head is
\begin{align}
    L_{\rm h}&=\Sigma_{\rm j}\rho_{\rm disk}\beta^3_{\rm h}c^3\nonumber\\
    &\simeq2.7\times10^{45} {\rm erg \ s^{-1}}\left(\frac{\Sigma_{\rm j}}{10^{24} {\rm cm^2}}\right)\nonumber\\
    &\left(\frac{\rho_{\rm disk}}{10^{-10}{\rm g \ cm^{-3}}}\right)\left(\frac{\beta_{\rm h}}{1}\right)^3,
\end{align}
where the detailed formulas for $\Tilde{L}$ and jet cross section $\Sigma_{\rm j}$ can be found in \cite{bromberg2011propagation}. The energy dissipated in cocoons (before the jet head breakout) is
\begin{equation}
\label{eq_inter}
    E_{\rm cd}=E_{\rm cj}=\frac{1}{2}\int_{0}^{t}{L_{\rm j}(1-\beta_{\rm h})dt}
\end{equation}

Once the jet head breaks out, the disk cocoon and jet cocoon, with internal energies $E_{\rm cd}$ and $E_{\rm cj}$, respectively, will expand and undergo adiabatic cooling at constant velocities until their respective breakouts. We calculate the subsequent dynamical evolution of the cocoons following the methodologies outlined in \cite{chen2024electromagnetic} and \cite{bromberg2011propagation}.

\section{Electromagnetic Emissions of Four Components}

For the four components that produced the remnant BH from the BBH merger, when the fast gas encounters the disk, a shock will form and sweep through the disk gas.
The optical depth above the shock is
\begin{equation}
    \tau_{\rm m}=\int_{R_{\rm s}}^{\infty}{\kappa\rho_{\rm disk}(r,R)dR},
\end{equation}
where opacity $\kappa=0.4~\rm cm^2g^{-1}$  which is appropriate for electron scattering for pure hydrogen matter~\citep{arnett1996supernovae,chevalier2011shock}, the photosphere radius $R_{\rm ph}$ can be obtained by setting optical depth equal to 1,
\begin{equation}
    \int_{R_{\rm ph}}^{\infty}{\kappa\rho_{\rm disk}(r,R)dR}=1,
\end{equation}

\begin{figure*}
\caption{The contour plots are the parameter space of $\eta_{\rm j}$($\eta_{\rm w}$) and $f_{\rm r}$, where the color regions are the contour levels for the g band peak flux $F_{\rm p,g}$, in units of $10^{-27} \rm erg/cm^2/s/Hz$. The blue, orange, and black lines with different linestyles are the contour levels for the delay, rising, and end time, i.e, $t_{\rm delay}$, $t_{\rm d, bo}$ and $t_{\rm end}$, in units of days. The allowed areas of all components for the GW190521 case are roughly marked as red stars. }
\includegraphics[width=0.5\linewidth,height=0.4\linewidth]{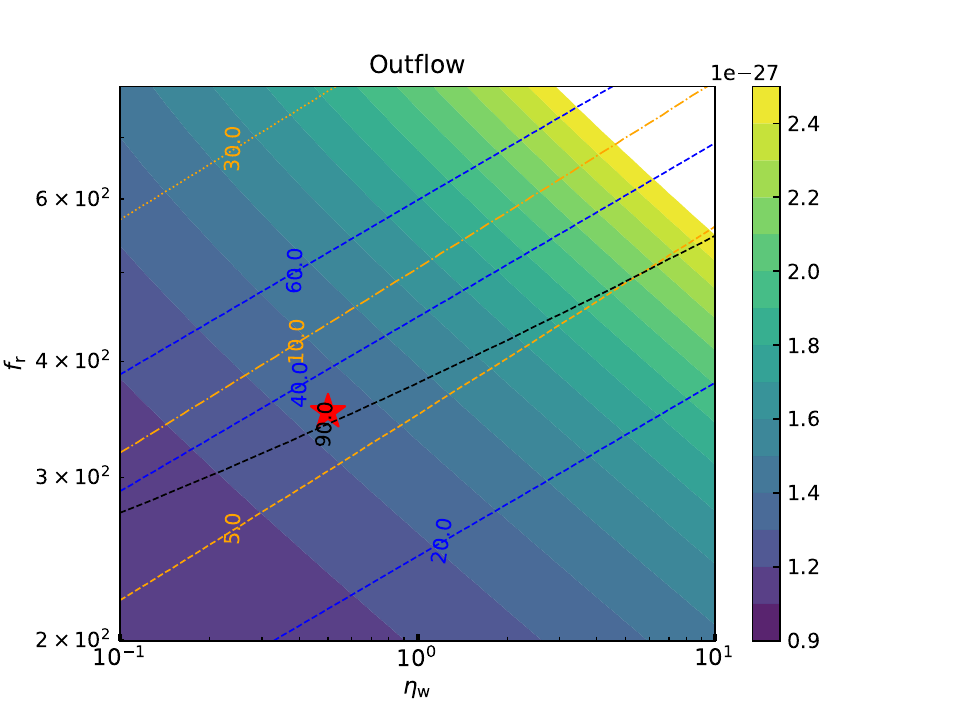}
\includegraphics[width=0.5\linewidth,height=0.4\linewidth]{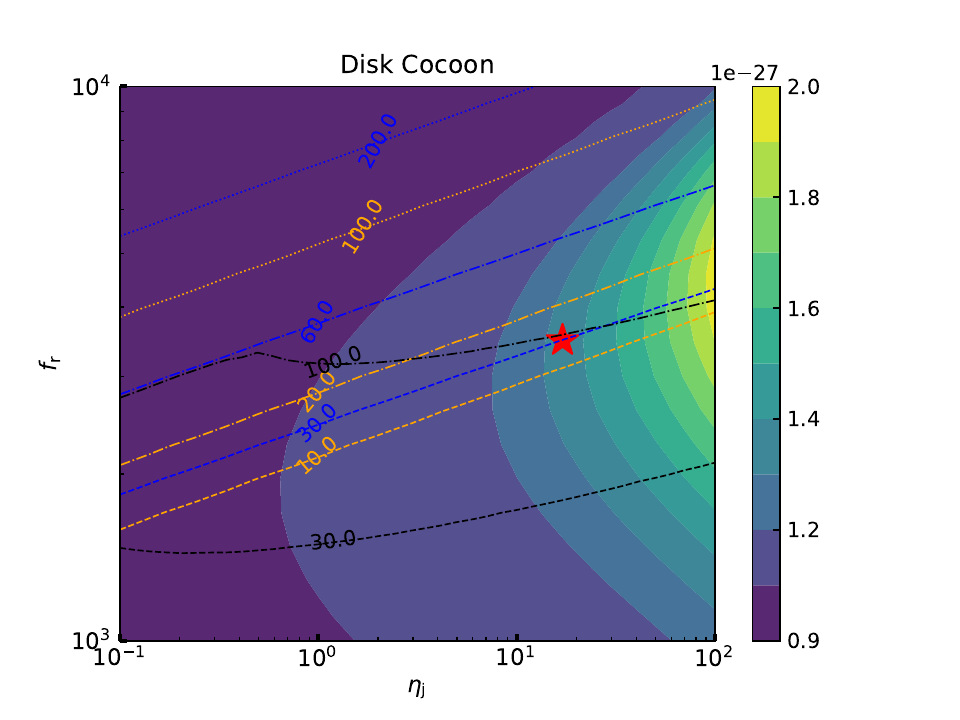}
\includegraphics[width=0.5\linewidth,height=0.4\linewidth]{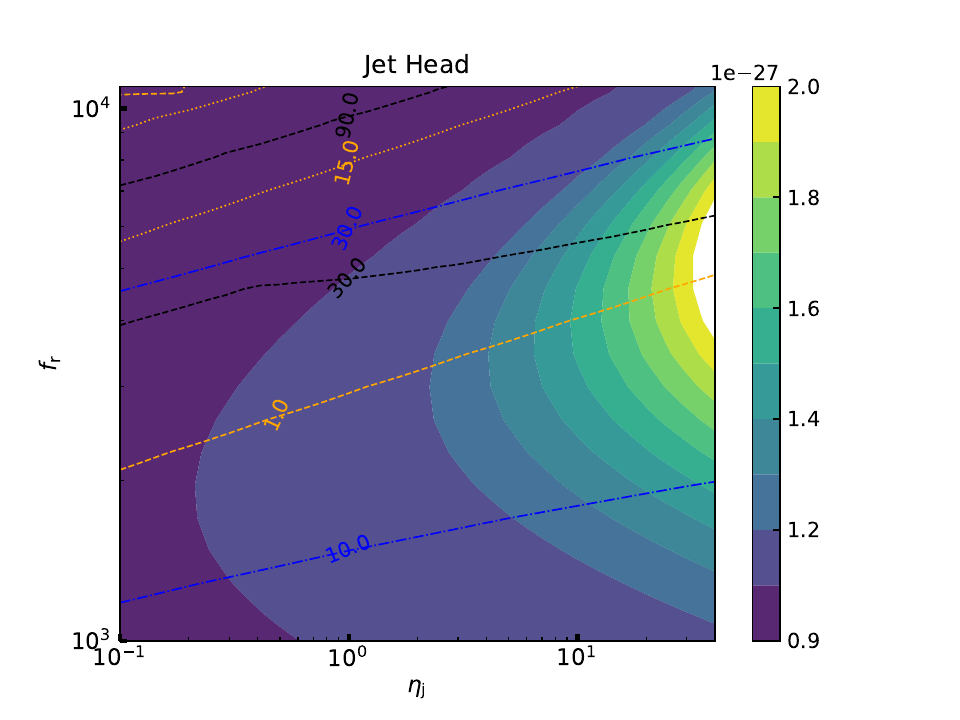}
\includegraphics[width=0.5\linewidth,height=0.4\linewidth]{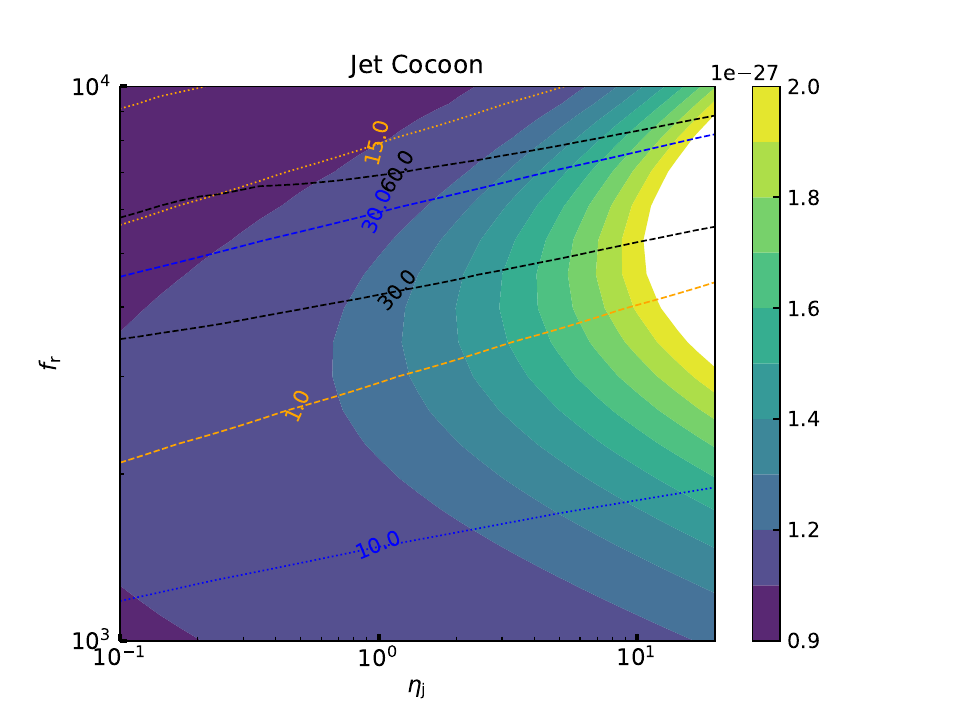}
\label{fig_con}
\end{figure*}

In the early time, the photon diffusion timescale $t_{\rm diff}$ is longer than the shock dynamic timescale $t_{\rm dyn}$, thus the radiation energy carried by photons is kept in the gas until $t_{\rm diff}\simeq t_{\rm dyn}$, this breakout condition is equivalent to $\tau_{\rm m}\simeq \tau_{\rm c}=c/v_{\rm s}$. The location satisfying the breakout condition is defined as the breakout radius $R_{\rm bo}$, and the time when the shock reaches $R_{\rm bo}$ is defined as $t_{\rm bo}$. Similarly, the time when the shock reaches $\tau_{\rm m}=1$ is defined as $t_{\rm ph}$.

After the shock breakout, the photons start to leak out, forming the observed breakout emission. The thermal energy carried by the gas to $t_{\rm bo}$ is $E_{\rm th}$. After $t_{\rm bo}$, we assume that all the thermal energy is radiative and is released entirely within a timescale $t_{\rm d, bo}$, which is the photon diffusion timescale from $R_{\rm bo}$ to $R_{\rm ph}$ at $t_{\rm bo}$, can be written as~\citep{ginzburg2012superluminous,ginzburg2013light}

 \begin{equation}
\label{eq_tdff}
   t_{\rm d,bo}\simeq\int_{R_{\rm bo}}^{\rm R_{\rm ph}}{\frac{6\kappa\rho_{\rm disk}(R-R_{\rm bo})}{c}dR}. 
\end{equation}
Under such an assumption, we can estimate bolometric lightcurves for the breakout emission and subsequent cooling emission through~\citep{chatzopoulos2012generalized}
\begin{equation}
\label{eq_lc}
L_{\rm bol}(t) =
\begin{cases} 
    0, & \text{if } t < t_{\rm bo}, \\
    \\
    \frac{E_{\rm th}}{t_{\rm d, bo}}\left[1- {\rm exp}\left(-\frac{t-t_{\rm bo}}{t_{\rm d, bo}}\right)\right], & \text{if }  t_{\rm bo}\le t < t_{\rm bo}+t_{\rm d, bo}, \\
    \\
    \frac{E_{\rm th}}{t_{\rm d, bo}}{\rm exp}\left(-\frac{t-t_{\rm bo}}{t_{\rm d, bo}}\right)(e-1), & \text{if } t \geq t_{\rm bo}+t_{\rm d, bo}, 
\end{cases}
\end{equation}
where the peak time $ t_{\rm bo}+t_{\rm d, bo}\simeq t_{\rm bo}+t_{\rm dyn}\simeq t_{\rm ph}$.
The effective temperature can thus be written as
\begin{equation}
    T_{\rm eff}=\left(\frac{L_{\rm bol}}{4\pi R^2_{\rm ph}\sigma_{\rm sb}}\right)^{1/4},
\end{equation}
where $\sigma_{\rm sb}$ is Stefan-Boltzmann constant. With such temperature, we can obtain the observed black-body SED at every time.

For the outflow, $R_{\rm bo}$ and $t_{\rm bo}$ can be obtained from Eqs.(\ref{eq_mov}) to (\ref{eq_mt}), allowing us to determine $t_{\rm d, bo}$ and the thermal injection energy $E_{\rm th} = E_{\rm w, th}(t_{\rm bo})$. Similarly, for the jet head, $R_{\rm bo}$ and $t_{\rm bo}$ are derived from Eq.(\ref{eq_vh}), yielding $t_{\rm d, bo}$ and $E_{\rm th} = L_{\rm h} t_{\rm d, bo}$. For cocoons, $R_{\rm bo}$ and $t_{\rm bo}$ are governed by both jet head breakout and subsequent adiabatic expansion, the injection energy at breakout $E_{\rm th}$ being a reduction of that in Eq.~(\ref{eq_inter}).

\section{Application to GW190521} 

\begin{figure*}
\caption{The best fitting light curves in the case of GW190521 for outflow and disk cocoon components. \textit{First row}: The resulting light curves for outflow in the g and r band, where data points and baseline emissions from \cite{graham2020candidate}, the total emission (solid) is the sum of the model result (dot-dashed) and baseline emission (dashed). The zero point of time is the GW detection time (vertical dotted line). \textit{Second row}: The similar results for the disk cocoon component.}

\includegraphics[width=0.5\linewidth,height=0.4\linewidth]{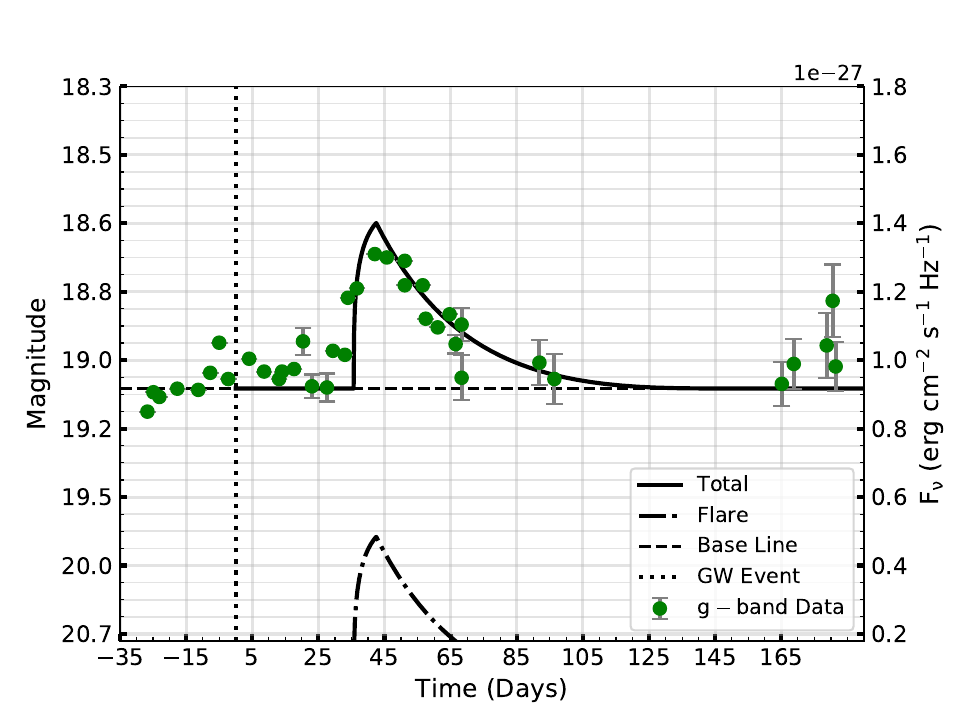}
\includegraphics[width=0.5\linewidth,height=0.4\linewidth]{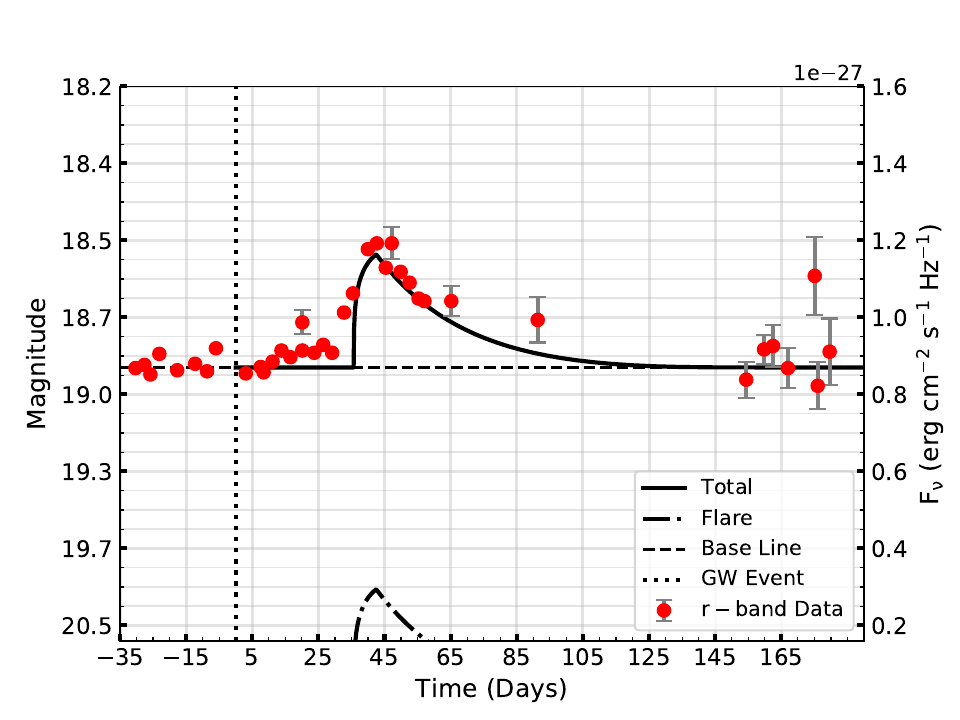}
\includegraphics[width=0.5\linewidth,height=0.4\linewidth]{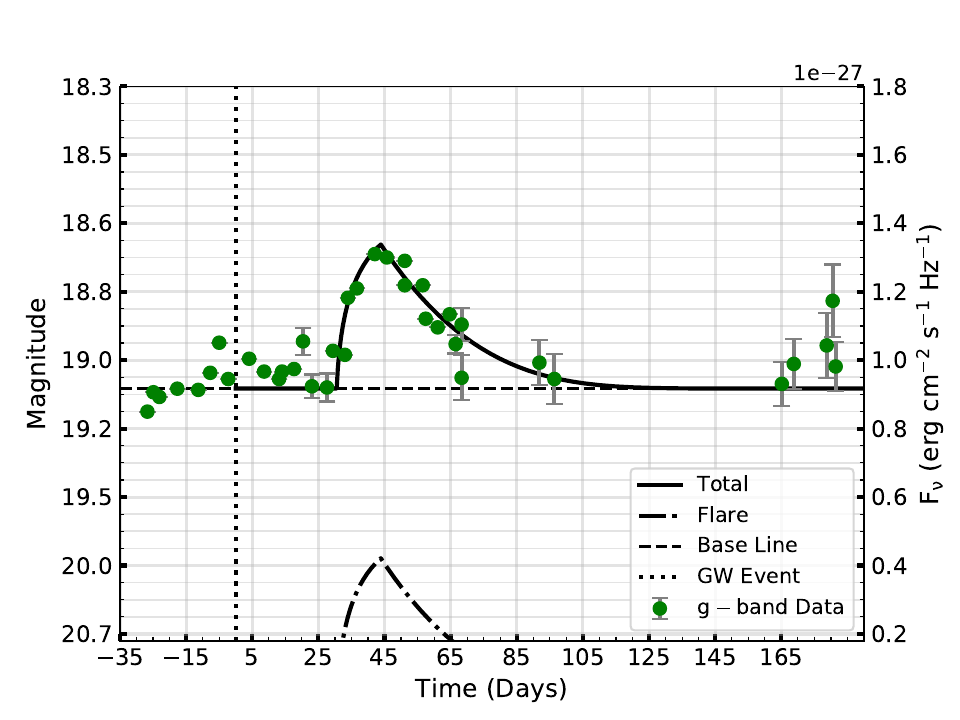}
\includegraphics[width=0.5\linewidth,height=0.4\linewidth]{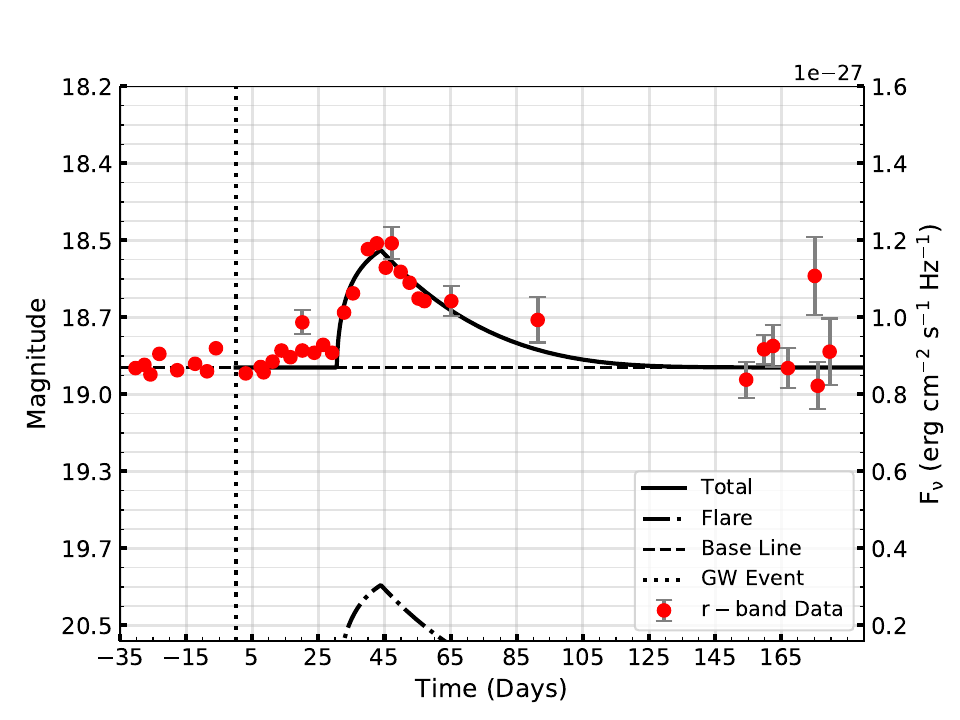}

\label{fig_res}
\end{figure*}

Firstly, we estimate the EM emissions in the case of GW190521 for the outflow component. With other parameters fixed as typical/observational values, we explore the allowed parameter space of $\eta_{w}$ and $f_{\rm r}$, highlighting the permitted region. The observed light curves offer constraints for the parameter space, which are: the delay time to GW detection $t_{\rm delay} \in [26, 36]$ days, the rising duration $t_{\rm d, bo} \in [7, 13]$ days, the end time for the declining phase (the time when flux drops to $9.5\times10^{-28} \ \text{erg/cm}^2/\text{s}/\text{Hz}$), $t_{\rm end} \in [90,100]$ days, and the peak flux of the g-band $F_{\rm p,g} \in [1.3, 1.4] \times 10^{-27} \ \text{erg/cm}^2/\text{s}/\text{Hz}$. These constraints are very stringent, resulting in a highly limited parameter space, with $\eta_{\rm w} \in [0.4, 0.7]$ and $f_{\rm r} \in [334, 378]$. The results are shown in Fig.~\ref{fig_con}.

Then, for the jet head, jet cocoon, and disk cocoon components, using the same method as in the outflow case, the allowed parameter spaces for $\eta_{\rm j}$ and $f_{\rm r}$ are explored as well. The results are presented in Fig.~\ref{fig_con}. For GW190521, only the disk cocoon has the allowed parameter space, i.e. $\eta_{\rm j} \in [13, 21]$ and $f_{\rm r} \in [3348, 3655]$, which is more extreme than the outflow case~\citep{2011MNRAS.418L..79T}.

\begin{figure*}
	\caption{The corresponding SEDs for the outflow (left) and disk cocoon (right) components in the case of GW190521. The dashed and solid line represents the SED at peak time ($t_{\rm delay} + t_{\rm d, bo}$) and at the moment just after the breakout ($t_{\rm delay} + 500$ s) respectively, where the baseline luminosity and frequency for the g and r band are also shown.}
	\includegraphics[width=0.5\linewidth,height=0.4\linewidth]{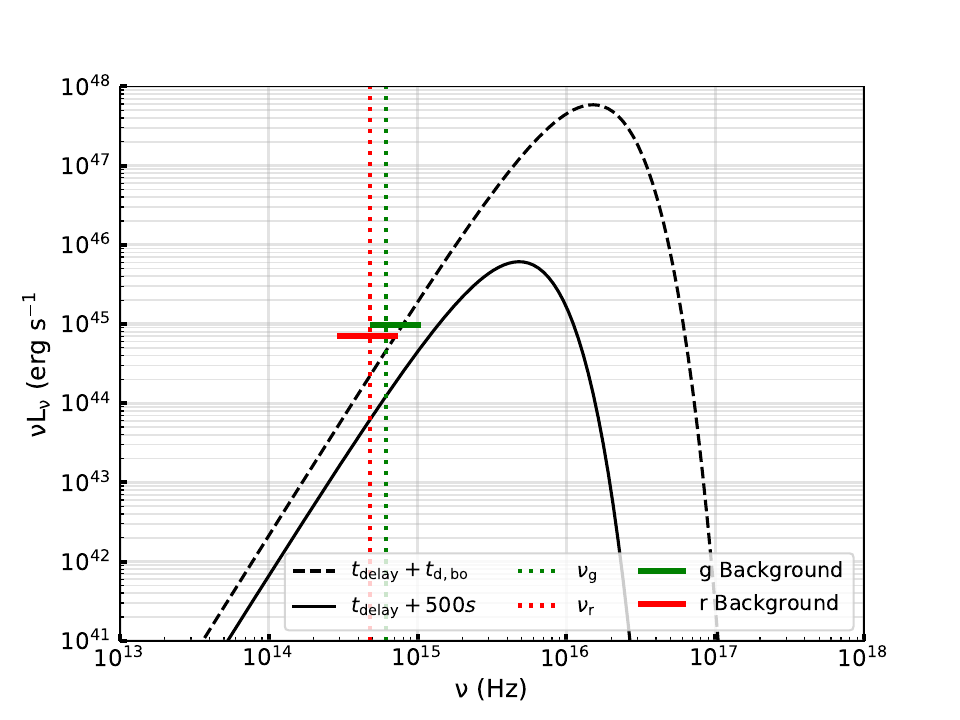}
	\includegraphics[width=0.5\linewidth,height=0.4\linewidth]{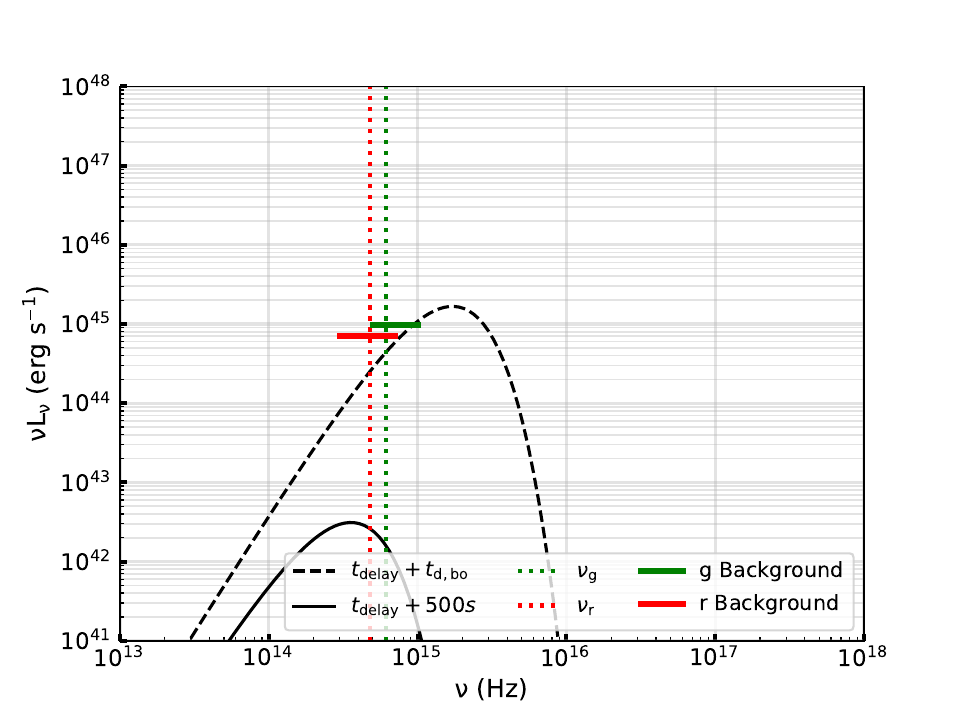}
	\label{fig_sed}
\end{figure*}

\begin{table*}
	\caption{The fitting parameters for outflow and disk cocoon in the case of GW190521. $M_{\rm SMBH}$, $M_{\rm bh}$ and $v_{\rm k}$ are obtained from~\cite{graham2020candidate}. The rest of the parameters are defined in the text.
	}
	\centering
	\begin{tabular}{@{}llllllllllll@{}}
		\hline\hline 
		{Outflow }	&	$M_{\rm SMBH}$	&	$M_{\rm bh}$	&	$v_{\rm w}$	&	$v_{\rm k}$	&$f_{\rm r}$		&	$\eta_{\rm w}$\\
		
		& [$M_{\odot}$]&  [$M_{\odot}$] & [cm/s] &  [cm/s] &[$R_{\rm Sch}$]  &   & \\
		\hline
		
		Value	&	$1\times10^8$	& 150		&$1\times10^{10}$	&$3\times10^{7}$	&	376	&	 0.63	\\
		\hline
		\\
		\hline\hline 
		{Disk cocoon }	&	$M_{\rm SMBH}$	&	$M_{\rm bh}$	&	$v_{\rm j}$	&	$v_{\rm k}$	&$f_{\rm r}$		&	$\eta_{\rm j}$\\
		
		& [$M_{\odot}$]&  [$M_{\odot}$] & [cm/s] &  [cm/s] &[$R_{\rm Sch}$]  &   & \\
		\hline
		
		Value	&	$1\times10^8$	& 150		&$3\times10^{10}$	&$3\times10^{7}$	&	3500	&	 16	\\
		\hline
	\end{tabular}
	\label{tab_para}
\end{table*}

For the outflow component, our best-fit results to the observed data are shown in Fig.~\ref{fig_res} (first row), as well as the corresponding SEDs in Fig.~\ref{fig_sed}, whose parameters are listed in Table~\ref{tab_para}. 
The zero point on the time axis is set as the GW detected time, denoted as the vertical dotted line. The total emission (solid) is the sum of the model result (dot-dashed) and the baseline emission (dashed). The resulting light curves rise on the 35th day after GW detection, this delay time is due to the outflow formation time $t_{\rm BHL}\simeq9~\rm days$ and outflow breakout time $t_{\rm bo} \simeq 26~\rm days$, i.e.
\begin{equation}
    t_{\rm delay}=t_{\rm BHL}+t_{\rm bo}\simeq35~\rm days,
\end{equation}
where $t_{\rm BHL}$ can be estimated from accretion timescale $t_{\rm BHL}\simeq Gm_{\rm bh}/v^3_{\rm k}$. Note that we ignore the time required for the remnant BH to cross the cavity created during the post-merger phase~\citep{kimura2021outflow,chen2024electromagnetic}. This timescale, denoted as $t_{\rm k}$, can be estimated by $t_{\rm k}=r_{\rm cav}/v_{\rm k}\simeq r_{\rm hill}/v_{\rm k}=(M_{\rm bh}/3M_{\rm SMBH})^{1/3}/v_{\rm k}$, where $r_{\rm hill}$ is the Hill radius~\citep{kimura2021outflow}. $t_{\rm k} $ usually has the magnitude of tens of days; however, $t_{\rm k}$ is highly uncertain because the remnant BH may not begin its movement from the center of the cavity. Therefore, we neglect it in this work. The rising phase of our light curve lasts for approximately $\sim 7$ days, which corresponds to the photon diffusion time $t_{\rm d, bo}$, where the peak bolometric luminosity $L_{\rm bol,p}\sim6\times10^{47}$ erg/s. The total injected energy $E_{\rm th} \sim 10^{53}$ erg is high. We show the SED at the peak time ($t_{\rm delay} + t_{\rm d, bo}$) and just after the breakout ($t_{\rm delay} + 500$ s) in Figure~\ref{fig_sed}. We find that the peak frequency is always in the range of $3\times10^{15}$ Hz to $2\times10^{16}$ Hz, within the UV band.

For the disk cocoon, the best-fitting light curves and corresponding SEDs are shown in Fig.~\ref{fig_res} and Fig.~\ref{fig_sed}, with parameters listed in Table~\ref{tab_para}. The delay time is approximately $t_{\rm delay} \sim 30$ days, with a rising phase lasting around 12 days. The peak bolometric luminosity is $L_{\rm bol,p} \sim 2\times10^{45}$ erg/s, and the total injected energy is $4 \times 10^{51}$ erg. The SED peaks at $4 \times 10^{14}$ to $2 \times 10^{15}$ Hz, ranging from optical to UV bands, which is lower than in the outflow case.

\section{Discussions and Conclusions}
\label{sec_disc}
In this study, we investigate the shock breakout and subsequent cooling emissions from the outflow, jet head, jet cocoon, and disk cocoon powered by remnant BHs from BBH mergers in the AGN disk. To explain the observed light curve of GW190521, we explored the parameter space for each component, based on previous dynamic studies and observations. Our findings indicate that only the outflow or the disk cocoon could be more likely responsible for the observations. We present the best-fit light curves and corresponding SEDs, which peak in the UV band for the outflow and in the optical-UV range for the disk cocoon. The injected energy is approximately $10^{53}$ erg for the outflow and $10^{51}$ erg for the disk cocoon.

Although this work focuses on GW190521, our methods for constraining parameter spaces and calculating light curves are broadly applicable. With additional observations of BBH gravitational wave detections and corresponding electromagnetic signals, our approach can be extended to further studies, enhancing our understanding of the underlying mechanisms and their governing principles.

\bibliography{reference}{}

\begin{thebibliography}{}
\expandafter\ifx\csname natexlab\endcsname\relax\def\natexlab#1{#1}\fi
\providecommand{\url}[1]{\href{#1}{#1}}
\providecommand{\dodoi}[1]{doi:~\href{http://doi.org/#1}{\nolinkurl{#1}}}
\providecommand{\doeprint}[1]{\href{http://ascl.net/#1}{\nolinkurl{http://ascl.net/#1}}}
\providecommand{\doarXiv}[1]{\href{https://arxiv.org/abs/#1}{\nolinkurl{https://arxiv.org/abs/#1}}}

\bibitem[{Arnett(1996)}]{arnett1996supernovae}
Arnett, D. 1996, Supernovae and nucleosynthesis: an investigation of the
  history of matter, from the big bang to the present, Vol.~7 (Princeton
  University Press)

\bibitem[{Artymowicz {et~al.}(1993)Artymowicz, Lin, \&
  Wampler}]{artymowicz1993star}
Artymowicz, P., Lin, D., \& Wampler, E. 1993, Astrophysical Journal, Part 1
  (ISSN 0004-637X), vol. 409, no. 2, p. 592-603., 409, 592

\bibitem[{Ashton {et~al.}(2021)Ashton, Ackley, Hernandez, \&
  Piotrzkowski}]{ashton2021current}
Ashton, G., Ackley, K., Hernandez, I.~M., \& Piotrzkowski, B. 2021, Classical
  and Quantum Gravity, 38, 235004

\bibitem[{Bartos {et~al.}(2017)Bartos, Kocsis, Haiman, \&
  M{\'a}rka}]{bartos2017rapid}
Bartos, I., Kocsis, B., Haiman, Z., \& M{\'a}rka, S. 2017, The Astrophysical
  Journal, 835, 165

\bibitem[{{Blustin} {et~al.}(2005){Blustin}, {Page}, {Fuerst},
  {Branduardi-Raymont}, \& {Ashton}}]{2005A&A...431..111B}
{Blustin}, A.~J., {Page}, M.~J., {Fuerst}, S.~V., {Branduardi-Raymont}, G., \&
  {Ashton}, C.~E. 2005, \aap, 431, 111, \dodoi{10.1051/0004-6361:20041775}

\bibitem[{Bondi(1952)}]{bondi1952spherically}
Bondi, H. 1952, Monthly Notices of the Royal Astronomical Society, 112, 195

\bibitem[{Bromberg {et~al.}(2011)Bromberg, Nakar, Piran,
  {et~al.}}]{bromberg2011propagation}
Bromberg, O., Nakar, E., Piran, T., {et~al.} 2011, The Astrophysical Journal,
  740, 100

\bibitem[{Campanelli {et~al.}(2007)Campanelli, Lousto, Zlochower, \&
  Merritt}]{campanelli2007maximum}
Campanelli, M., Lousto, C.~O., Zlochower, Y., \& Merritt, D. 2007, Physical
  Review Letters, 98, 231102

\bibitem[{Cantiello {et~al.}(2021)Cantiello, Jermyn, \&
  Lin}]{cantiello2021stellar}
Cantiello, M., Jermyn, A.~S., \& Lin, D.~N. 2021, The Astrophysical Journal,
  910, 94

\bibitem[{Chatzopoulos {et~al.}(2012)Chatzopoulos, Wheeler, \&
  Vinko}]{chatzopoulos2012generalized}
Chatzopoulos, E., Wheeler, J.~C., \& Vinko, J. 2012, The Astrophysical Journal,
  746, 121

\bibitem[{Chen \& Dai(2024)}]{chen2024electromagnetic}
Chen, K., \& Dai, Z.-G. 2024, The Astrophysical Journal, 961, 206

\bibitem[{Chen {et~al.}(2023)Chen, Ren, \& Dai}]{chen2023role}
Chen, K., Ren, J., \& Dai, Z.-G. 2023, The Astrophysical Journal, 948, 136

\bibitem[{Cheng \& Wang(1999)}]{cheng1999formation}
Cheng, K., \& Wang, J.-M. 1999, The Astrophysical Journal, 521, 502

\bibitem[{Chevalier \& Irwin(2011)}]{chevalier2011shock}
Chevalier, R.~A., \& Irwin, C.~M. 2011, The Astrophysical Journal Letters, 729,
  L6

\bibitem[{{Collin} \& {Zahn}(1999)}]{collin1999star}
{Collin}, S., \& {Zahn}, J.-P. 1999, \apss, 265, 501,
  \dodoi{10.1023/A:1002191506811}

\bibitem[{DeLaurentiis {et~al.}(2023)DeLaurentiis, Epstein-Martin, \&
  Haiman}]{delaurentiis2023gas}
DeLaurentiis, S., Epstein-Martin, M., \& Haiman, Z. 2023, Monthly Notices of
  the Royal Astronomical Society, 523, 1126

\bibitem[{Dermer \& Menon(2009)}]{dermer2009high}
Dermer, C.~D., \& Menon, G. 2009, High energy radiation from black holes: gamma
  rays, cosmic rays, and neutrinos, Vol.~17 (Princeton University Press)

\bibitem[{Dexter \& Begelman(2023)}]{10.1093/mnrasl/slad182}
Dexter, J., \& Begelman, M.~C. 2023, Monthly Notices of the Royal Astronomical
  Society: Letters, 528, L157, \dodoi{10.1093/mnrasl/slad182}

\bibitem[{Edgar(2004)}]{edgar2004review}
Edgar, R. 2004, New Astronomy Reviews, 48, 843

\bibitem[{Faucher-Gigu{\`e}re \& Quataert(2012)}]{faucher2012physics}
Faucher-Gigu{\`e}re, C.-A., \& Quataert, E. 2012, Monthly Notices of the Royal
  Astronomical Society, 425, 605

\bibitem[{{Fender}(2002)}]{2002LNP...589..101F}
{Fender}, R. 2002, in Relativistic Flows in Astrophysics, ed. A.~W. {Guthmann},
  M.~{Georganopoulos}, A.~{Marcowith}, \& K.~{Manolakou}, Vol. 589, 101,
  \dodoi{10.48550/arXiv.astro-ph/0109502}

\bibitem[{Ginzburg \& Balberg(2012)}]{ginzburg2012superluminous}
Ginzburg, S., \& Balberg, S. 2012, The Astrophysical Journal, 757, 178

\bibitem[{Ginzburg \& Balberg(2013)}]{ginzburg2013light}
---. 2013, The Astrophysical Journal, 780, 18

\bibitem[{Gonzalez {et~al.}(2007)Gonzalez, Sperhake, Bruegmann, Hannam, \&
  Husa}]{gonzalez2007maximum}
Gonzalez, J.~A., Sperhake, U., Bruegmann, B., Hannam, M., \& Husa, S. 2007,
  Physical Review Letters, 98, 091101

\bibitem[{Graham {et~al.}(2020)Graham, Ford, McKernan, Ross, Stern, Burdge,
  Coughlin, Djorgovski, Drake, Duev, {et~al.}}]{graham2020candidate}
Graham, M., Ford, K., McKernan, B., {et~al.} 2020, Physical review letters,
  124, 251102

\bibitem[{Graham {et~al.}(2023)Graham, McKernan, Ford, Stern, Djorgovski,
  Coughlin, Burdge, Bellm, Helou, Mahabal, {et~al.}}]{graham2023light}
Graham, M.~J., McKernan, B., Ford, K.~S., {et~al.} 2023, The Astrophysical
  Journal, 942, 99

\bibitem[{Herrmann {et~al.}(2007)Herrmann, Hinder, Shoemaker, \&
  Laguna}]{herrmann2007unequal}
Herrmann, F., Hinder, I., Shoemaker, D., \& Laguna, P. 2007, Classical and
  Quantum Gravity, 24, S33

\bibitem[{Hoyle \& Lyttleton(1939)}]{hoyle1939effect}
Hoyle, F., \& Lyttleton, R.~A. 1939in , Cambridge University Press, 405--415

\bibitem[{Kato {et~al.}(1998)Kato, Fukue, \& Mineshige}]{kato1998black}
Kato, S., Fukue, J., \& Mineshige, S. 1998, Black-hole accretion disks

\bibitem[{Kimura {et~al.}(2021)Kimura, Murase, \& Bartos}]{kimura2021outflow}
Kimura, S.~S., Murase, K., \& Bartos, I. 2021, The Astrophysical Journal, 916,
  111

\bibitem[{Li {et~al.}(2022)Li, Lai, \& Rodet}]{li2022long}
Li, J., Lai, D., \& Rodet, L. 2022, The Astrophysical Journal, 934, 154

\bibitem[{Liu {et~al.}(2018)Liu, Murase, Inoue, Ge, \& Wang}]{liu2018can}
Liu, R.-Y., Murase, K., Inoue, S., Ge, C., \& Wang, X.-Y. 2018, The
  Astrophysical Journal, 858, 9

\bibitem[{Ma \& Wang(2024)}]{ma2024high}
Ma, Z.-P., \& Wang, K. 2024, The Astrophysical Journal, 970, 127

\bibitem[{Morton {et~al.}(2023)Morton, Rinaldi, Torres-Orjuela, Derdzinski,
  Vaccaro, \& Del~Pozzo}]{morton2023gw190521}
Morton, S.~L., Rinaldi, S., Torres-Orjuela, A., {et~al.} 2023, Physical Review
  D, 108, 123039

\bibitem[{Murase(2024)}]{murase2024interacting}
Murase, K. 2024, Physical Review D, 109, 103020

\bibitem[{{Pounds} {et~al.}(2003){Pounds}, {Reeves}, {King}, {Page}, {O'Brien},
  \& {Turner}}]{2003MNRAS.345..705P}
{Pounds}, K.~A., {Reeves}, J.~N., {King}, A.~R., {et~al.} 2003, \mnras, 345,
  705, \dodoi{10.1046/j.1365-8711.2003.07006.x}

\bibitem[{Rodr{\'\i}guez-Ram{\'\i}rez
  {et~al.}(2024)Rodr{\'\i}guez-Ram{\'\i}rez, Bom, Fraga, \&
  Nemmen}]{rodriguez2024optical}
Rodr{\'\i}guez-Ram{\'\i}rez, J., Bom, C., Fraga, B., \& Nemmen, R. 2024,
  Monthly Notices of the Royal Astronomical Society, 527, 6076

\bibitem[{Stone {et~al.}(2017)Stone, Metzger, \& Haiman}]{stone2017assisted}
Stone, N.~C., Metzger, B.~D., \& Haiman, Z. 2017, Monthly Notices of the Royal
  Astronomical Society, 464, 946

\bibitem[{Tagawa {et~al.}(2020)Tagawa, Haiman, \& Kocsis}]{tagawa2020formation}
Tagawa, H., Haiman, Z., \& Kocsis, B. 2020, The Astrophysical Journal, 898, 25

\bibitem[{Tagawa {et~al.}(2024)Tagawa, Kimura, Haiman, Perna, \&
  Bartos}]{tagawa2024shock}
Tagawa, H., Kimura, S.~S., Haiman, Z., Perna, R., \& Bartos, I. 2024, The
  Astrophysical Journal, 966, 21

\bibitem[{{Tchekhovskoy} {et~al.}(2011){Tchekhovskoy}, {Narayan}, \&
  {McKinney}}]{2011MNRAS.418L..79T}
{Tchekhovskoy}, A., {Narayan}, R., \& {McKinney}, J.~C. 2011, \mnras, 418, L79,
  \dodoi{10.1111/j.1745-3933.2011.01147.x}

\bibitem[{Wang {et~al.}(2021{\natexlab{a}})Wang, Liu, Ho, \&
  Du}]{wang2021accretiona}
Wang, J.-M., Liu, J.-R., Ho, L.~C., \& Du, P. 2021{\natexlab{a}}, The
  Astrophysical Journal Letters, 911, L14

\bibitem[{Wang {et~al.}(2021{\natexlab{b}})Wang, Liu, Ho, Li, \&
  Du}]{wang2021accretionb}
Wang, J.-M., Liu, J.-R., Ho, L.~C., Li, Y.-R., \& Du, P. 2021{\natexlab{b}},
  The Astrophysical Journal Letters, 916, L17

\bibitem[{{Wang} {et~al.}(2023){Wang}, {Liu}, {Li}, {Songsheng}, {Yuan}, \&
  {Ho}}]{2023ApJ...958L..40W}
{Wang}, J.-M., {Liu}, J.-R., {Li}, Y.-R., {et~al.} 2023, \apjl, 958, L40,
  \dodoi{10.3847/2041-8213/ad0bd9}

\bibitem[{Wang {et~al.}(2019)Wang, Huang, \& Li}]{wang2019transient}
Wang, K., Huang, T.-Q., \& Li, Z. 2019, The Astrophysical Journal, 872, 157

\bibitem[{Weaver {et~al.}(1977)Weaver, McCray, Castor, Shapiro, \&
  Moore}]{weaver1977interstellar}
Weaver, R., McCray, R., Castor, J., Shapiro, P., \& Moore, R. 1977,
  Astrophysical Journal, Part 1, vol. 218, Dec. 1, 1977, p. 377-395., 218, 377

\bibitem[{Zhou \& Wang(2023)}]{zhou2023high}
Zhou, Z.-H., \& Wang, K. 2023, The Astrophysical Journal Letters, 958, L12

\bibitem[{Zhu {et~al.}(2021)Zhu, Wang, Zhang, Yang, Yu, \& Gao}]{zhu2021high}
Zhu, J.-P., Wang, K., Zhang, B., {et~al.} 2021, The Astrophysical Journal
  Letters, 911, L19

\end{thebibliography}
\bibliographystyle{aasjournal}

\end{document}